\documentstyle[12pt]{article}
\begin{document}

\setcounter{page}{0}
\thispagestyle{empty}

\vspace*{-1in}
\begin{flushright}CUPP-95/4\\
\end{flushright}
\vskip 50pt
\begin{center}
{\large\bf{ACCELERATOR, REACTOR, SOLAR AND }\\
\large\bf {ATMOSPHERIC NEUTRINO OSCILLATION:}\\
\large\bf{BEYOND THREE GENERATIONS\\}}

\vskip 45pt
{\it Srubabati Goswami \\
Department of Pure Physics,\\University of Calcutta,\\
92 Acharya Prafulla Chandra Road,\\Calcutta 700 009, INDIA.}\\

\vskip 15pt

P.A.C.S. Nos.: 14.60.Pq, 14.60.Lm, 96.40.Tv, 96.60 Kx

\vskip 10pt

\end{center}

\vskip 15pt

We perform a phenomenological analysis of neutrino oscillation
in a four generation framework introducing an additional sterile
neutrino. In such a scenario, more than one pattern is possible that
can accommodate three hieararchically different mass squared differences
as required by the  present experiments. We considered
two different spectrums. Choosing the ${\Delta{m}}^2$s
in the ranges suitable for the LSND, atmospheric and solar neutrino
oscillation, limits on the mixing angles are derived, consistent with
the most restrictive accelerator and reactor data as well as the
atmospheric and solar neutrino results. The allowed mixing angles are
found to be constrained very severely in both cases.
For one mass pattern in the combined allowed zone the
atmospheric anomaly can be explained by  $\nu_e - \nu_{\mu}$
oscillation whereas for the other the $\nu_{\mu} - \nu_{\tau}$ channel
is preferred. The accelerator experiments CHORUS and NOMAD have
different sensitivities in these regions and they can distinguish
between the two choices.

\vskip 15pt
\parindent 0pt
June 28, 1995

\parindent 30pt

\newpage

\section {\bf INTRODUCTION}
The question whether neutrinos have a non-zero mass or not has remained
one of the most tantalising issues in present day  physics.
In the standard model of electroweak theory the neutrinos are
considered to be massless. But there is no compelling theoretical
reason behind this assumption. Most extensions of the standard model
allow small but non-zero neutrino mass.
A way for probing small neutrino masses and the mixing between
different neutrino flavours is provided by neutrino oscillations.
Considering only two generations for simplicity, the probability
that an initial $\nu_{\alpha}$ of energy $E$ gets converted to a
$\nu_{\beta}$ after travelling a distance $L$ in vacuum is
\begin{equation}
P_{\nu_{\alpha}\nu_{\beta}} = \sin{^2}2\theta \sin{^{2}}(1.27{\Delta
m}^{2}L/E)
\label{pnu}
\end{equation}
where $\theta$ is the mixing angle in vacuum. ${\Delta m}^{2}$ denotes
the mass difference squared in $eV^{2}$. $L/E$ is in meter/MeV. The
oscillatory character is embedded in the second factor in
(\ref{pnu}). The detection of this phenomenon in an experiment
requires $E/L \simeq {\Delta m}^{2}$.

Recently the Liquid Scintillator Neutrino Detector (LSND)
collaboration has declared its results for a
positive evidence of $\overline{\nu}_{\mu} - \overline{\nu}_e$
oscillation \cite{lsnd}. Prior to this, indications of neutrino
oscillations came from the well known solar neutrino problem and the
atmospheric neutrino anomaly.

If indeed neutrino flavour oscillation takes place, the most
sensitive value of ${\Delta m}^{2}$ for detecting such a phenomenon in
LSND is $\sim$ 6eV$^2$, which is in the right ballpark for the cold
plus hot dark matter scenario for structure formation in the early
universe \cite{primack}. It remains to be seen whether the LSND results
stand the test of time but already this has added a new
impetus to the issue of neutrino mass and mixing and a number of
investigations have been carried out recently, discussing the possible
impact of this on particle physics, astrophysics and cosmology
\cite{silk,new}.

The observed suppression of the solar neutrino fluxes as compared to
the theoretical predictions constitutes the long standing solar
neutrino problem.
A purely astrophysical solution to this, attributing
the deficit to an inaccurate prediction of the fluxes by the standard
solar models \cite{ssm} is disfavoured by the present data
\cite{langacker1}. If neutrinos are massive, a plausible explanation
to the solar neutrino problem is neutrino oscillation in vacuum
\cite{vac} or the Mikheyev-Smirnov-Wolfenstein (MSW) \cite{msw} effect
of matter enhanced oscillations. The basic idea is flavour conversion
of $\nu_{e}$ to another species --  active or  sterile -- to which the
detector is not sensitive.
The two generation oscillation explanation for
the solar neutrino problem requires ${{\Delta}m}^2 \sim 6 \times
10^{-6} eV^2$ and $\sin^{2}2\theta \sim 7 \times 10^{-3}$ (non-adiabatic
solution) and ${{\Delta}m}^2 \sim 9 \times 10^{-6} eV^2$ and
$\sin^{2}2\theta \sim 0.6$ (large mixing angle solution)
\cite{langacker2} for the MSW matter oscillation to an  active neutrino.
If instead one considers oscillation to  sterile neutrinos as a possible
solution, the large angle region is excluded at 99\% C.L. \cite{langacker2}.
This region is also not consistent with the bound on the number of
neutrino species in big-bang nucleosynthesis \cite{shi}.
Oscillation in vacuum to an active neutrino requires
${{\Delta}m}^2 \sim (0.45-1.2) \times 10^{-10} eV^2$ and
$\sin^{2}2\theta \sim (0.6-1.0)$ \cite{hata}. The sterile neutrino
alternative for this case is ruled out by the present data at 98\% C.L
\cite{krastev}.

The primary components of the cosmic-ray flux interact with the earths
atmosphere producing pions and kaons which can decay as:
\begin{center}
$\pi^{+}(K^{\pm}) \rightarrow \mu^{\pm} +
\nu_{\mu}(\overline{\nu}_{\mu})$
\end{center}
\begin{center}
$\mu^{\pm} \rightarrow e^{\pm} + \nu_e(\overline{\nu}_e) +
\overline{\nu}_{\mu}(\nu_{\mu})$
\end{center}
These neutrinos can be detected by imaging water
$\breve{C}$erenkov detectors, -- Kamiokande
\cite{hirata,fukuda} and IMB \cite{imb} -- or
using iron calorimeters as is done in  Fr\'{e}jus \cite{fr}, Nusex
\cite{nus} and Soudan2 \cite{soudan}.
To reduce the uncertainty in the absolute flux values the usual
practice is to present the  ratio of ratios $R$ \cite{fogli2} :
\begin{equation}
R = {\frac{(\nu_\mu + \overline{\nu}_{\mu})/
(\nu_e + \overline{\nu}_e)_{\rm obsvd}}
{(\nu_\mu + \overline{\nu}_{\mu})/(\nu_e + \overline{\nu}_e)_{\rm MC}}}
\label{ratm}
\end{equation}
where {MC} denotes the Monte-Carlo simulated ratio.
Kamiokande and IMB find $R$ to be less than the expected
value of unity. This deviation is known as the atmospheric neutrino
anomaly. The preliminary results from Soudan2
agrees with this but Fr\'{e}jus and Nusex do not support this
conclusion.
The atmospheric anomaly, if it exists, can be explained by either
$\nu_{\mu} - \nu_e$ or $\nu_{\mu} - \nu_{\tau}$ oscillations in a
two generation picture. The analysis of the new multi-GeV data
as well as the previous sub-GeV data of the
Kamiokande collaboration predicts the following best-fit
parameters (${\Delta m}^{2}, \sin^2 2\theta$) = (1.8 $\times 10^{-2}
eV^{2}, 1.0)$  for $\nu_{\mu} - \nu_e$ oscillation and
(1.6 $\times 10^{-2}eV^{2}, 1.0)$ for $\nu_{\mu} - \nu_{\tau}$
oscillation \cite{fukuda}. Since the required mixing angle is large,
oscillation to sterile neutrinos is again inconsistent with the
nucleosynthesis constraints \cite{shi}.

The three neutrino oscillation phenomena mentioned above
--  namely the solar neutrino problem, the atmospheric neutrino anomaly
and the $\overline{\nu}_{\mu} - \overline{\nu}_e$ oscillations observed by
LSND group --  require three hieararchically different mass ranges and
a simultaneous explanation of all of them would need at least four
generation of neutrinos \cite{silk}.
LEP data reveals that there are three light active neutrino
species. This is supported by the requirements of nucleosynthesis
in the early universe. So the fourth neutrino has to be sterile.
Introducing this new neutrino one can attempt separate two
generation treatments for each but a more comprehensive approach
would be to determine the parameter ranges consistent with all the
experiments by a combined four generation analysis which can reveal
the full implications of each experimental data on the other.
In this paper we perform such an oscillation analysis in a four family
picture.

In a four generation scenario there are six ${\Delta m}^2$s, three of
which are independent and six mixing angles neglecting the CP violating
phases. We assume a minimal four flavour mixing scheme, in which the
sterile neutrino mixes only with the electron neutrino, thus reducing
the number of mixing angles to four.
We do not make any assumptions regarding these mixing
angles allowing them to cover the whole range from 0 to $\pi/2$.
Guided by the present data on neutrino oscillation we consider
two different sets of hieararchical mass squared differences:\\
\noindent (i)  $\Delta_{13} \simeq \Delta_{23} \simeq
\Delta_{34} = \Delta_{LSND}$ in the LSND range,\\
$\Delta_{12} \simeq\Delta_{24} = \Delta_{ATM}$ as preferred by the
atmospheric neutrino data and \\
$\Delta_{14} = \Delta_{SOLAR}$ either in the MSW or in the vacuum
oscillation range.\\
\noindent (ii)$\Delta_{12} \simeq
\Delta_{13} \simeq \Delta_{24} \simeq \Delta_{34} = \Delta_{LSND}$ \\
$\Delta_{23} = \Delta_{ATM}$\\
$\Delta_{14} = \Delta_{SOLAR}$\\
$\Delta_{ij} = \mid{{m_j}^2 - {m_i}^2}\mid$.
The hieararchy in the absolute values of neutrino masses as implied by
the above spectrums can be classified in general as\\
\noindent{Case (i)} :
(a)$ m_{1}^2 \simeq m_{4}^2 << m_{2}^2 \approx m_{3}^2$~~ or ~~
(b)$ m_{1}^2 \simeq m_{4}^2 >> m_{2}^2 \approx m_{3}^2$ \\
\noindent{Case (ii)}:
(a)$m_{1}^2 \simeq m_{4}^2 \approx m_{2}^2 << m_{3}^2$  ~~or ~~
(b)$m_{1}^2 \simeq m_{4}^2 \approx m_{2}^2 >> m_{3}^2$  \\
These are shown schematically in fig. 1. The $\simeq$ and $\approx$
signs imply, differences by $\Delta_{SOLAR}$ and $\Delta_{ATM}$
are neglected respectively, while the $>>$ or $<<$ sign means
difference by $\Delta_{LSND}$.
Neutrino oscillation analysis remains the same for types (a) and (b) in
both cases and one has to invoke some other experimental constraints
like that obtained from neutrinoless double beta decay for a
distinction between these \cite{minakata}.

Models for neutrino masses and mixings assuming the existence of
sterile states besides the three active flavours have been
discussed before \cite{model1}. After the declaration of the
LSND results models involving  extra singlet neutrinos have been
constructed by many authors \cite{model2}. Our investigation is not
motivated by any particular model. Rather, we do a phenomenological
analysis fixing the ${\Delta m}^2$s in the ranges suitable for
explaining the solar neutrino, atmospheric neutrino and the LSND
results and determine the allowed values of the mixing angles, which
can reconcile all the experimental data. The perspectives of the accelerator
experiments CHORUS and NOMAD, searching for $\nu_{\mu} - \nu_{\tau}$
oscillations, are discussed in the light of our findings.

The plan of the paper is as follows.
In section 2 we summarize the experimental results relevant for our purpose.
In the following section the general multi-generation formalism for
studying neutrino oscillations is developed. In Section 4 we calculate the
survival and transition probabilities for the various experiments for
the mass patterns (i) and (ii). The results of our analysis
and some discussions are presented in section 5. We end in section 6
with a short summary and conclusions.

\section{Experimental Results}
\subsection{Laboratory experiments}
The laboratory experiments can serve as an
important tool for neutrino oscillation search. These are
either accelerator or reactor based and in general are of two types: \\
(i) Disappearance experiments: in which one looks for a reduction in
the initial neutrino flux due to oscillation to some other flavour to
which the detector is not sensitive. \\
(ii) Appearance experiments: in which one searches for a new neutrino
flavour, absent in the initial beam, which can arise from
oscillation.\\
Prior to the LSND results all the laboratory experiments were
consistent with no neutrino oscillations \cite{prog} and provided
exclusion regions in the
$\Delta{m}^2 - \sin^2{2\theta}$ plane. In the region of large
$\Delta{m}^2$, the $\sin{^{2}}(1.27{\Delta m}^{2}L/E)$ term
$\rightarrow$ 0.5 and $P_{{\nu_\alpha}{\nu_\beta}} = 0.5
\sin^2{2\theta}$. Thus the limits on $\sin^2{2\theta}$ from the
exclusion plots can be used to provide bounds on
$P_{\nu_{\alpha}\nu_{\beta}}$. Below we summarize the laboratory
experiments which give most
stringent bounds on $P_{\nu_{\alpha}\nu_{\beta}}$.

The reactor experiments are suitable for a search of
$P_{\overline{\nu}_e\overline{\nu}_e}$ using disappearance technique.
The most stringent constraint is provided by  Bugey which measures the
spectrum of $\overline{\nu}_e$, coming from the Pressurized Water Reactors
running at the Bugey nuclear power plant, at 15, 40, and 95 metre
using neutron detection techniques. The 90\% C.L. exclusion contour
implies $1 - P_{\overline{\nu}_e\overline{\nu}_e}
\stackrel{\textstyle{<}}{\sim} $ 0.05
\cite{bugey}.

Accelerators produce mainly $\nu_{\mu}$ beams -- the $\nu_e$ component
being small and poorly determined.
The most restrictive accelerator based disappearance
experiment measuring $P_{\overline{\nu}_{\mu}\overline{\nu}_{\mu}}$ is
CDHSW. The 90\%
C.L. bound one obtains from the exclusion plot presented by the CDHSW
collaboration is $P_{\nu_{\mu}\nu_{\mu}} \stackrel{\textstyle{>}}{\sim}
$ 0.95 \cite{cdhsw}.

Accelerator based appearance experiments searching for
$\overline{\nu}_{\mu} \rightarrow \overline{\nu}_e$ oscillations
are LSND, KARMEN and E776 at BNL.
The last two experiments are consistent with no neutrino
oscillation. KARMEN has so far quoted an upper limit on the
oscillation probability as $P_{\overline{\nu}_{\mu}\overline{\nu}_e}
\leq 3.1 \times 10^{-3}$ (90\% C.L.) \cite{karmen} whereas from the
two flavour exclusion areas presented by BNL one gets
$P_{\overline{\nu}_{\mu}\overline{\nu}_e}
\stackrel{\textstyle{<}}{\sim} 1.5  \times 10^{-3}$
(90\% C.L.) \cite{bnl} which is the more restrictive of the two.
LSND has claimed a  positive evidence for neutrino
oscillations and  reports an excess of
${16.4}^{+9.7}_{-8.9} \pm$ 3.3 events
over the estimated background which, if interpreted in terms of
neutrino oscillations, corresponds to a probability
$P_{\overline{\nu}_{\mu}\overline{\nu}_e}$ of
$({0.34}^{+0.20}_{-0.18} \pm 0.07)$\% \cite{lsnd}.

E531 at Fermilab looks for ${\nu_{\mu} - \nu_{\tau}}$ oscillation, by
$\nu_{\tau}$ appearance.
Their results can be translated into an upper
limit on the oscillation probability: $P_{\nu_{\mu}\nu_{\tau}}
\stackrel{\textstyle{<}}{\sim}$ 2 $\times 10^{-3}$ \cite{e531}.

CHORUS \cite{chorus} and NOMAD \cite{nomad} at CERN search for
$\nu_{\mu} - \nu_{\tau}$ oscillation. With the CERN SPS designed to
deliver 2.4 $\times 10^{19}$ protons, CHORUS and NOMAD are sensitive to
a minimum oscillation probability of $10^{-4}$.

These experiments are sensitive to different $\Delta{m}^2$, because the
$L/E$ factor is different for each. In Table 1 we summarize the
characteristics of the experiments which are most constraining and of
CHORUS and NOMAD whose results are awaited.

\subsection{Atmospheric neutrinos}
Among the experiments measuring the atmospheric neutrino flux, data of
most statistical
significance have been collected by the Kamiokande and the IMB
collaborations.
For neutrinos of energy less than $\sim$ 1 GeV, IMB finds $R$ = $0.54
\pm 0.05 \pm 0.12$ \cite{imb} in agreement with the Kamiokande data
$R$ = $0.60^{+0.06}_{-0.05} \pm 0.05$ in this energy range
\cite{hirata,fukuda}. Recently the Kamiokande group has
published the results of the measurement of the flux ratio in the
multi-GeV energy range \cite{fukuda}. They found $R$ =
$0.57^{+0.08}_{-0.07} \pm 0.07$ in good agreement with the sub-GeV
value. Another aspect of this
measurement that can independently point towards neutrino
oscillation is the dependence of $R$ on the zenith-angle. The
multi-GeV Kamiokande data reveals a dependence on the zenith-angle
unlike the sub-GeV data, though the statistical significance of
this result has been questioned \cite{cern}. For the purpose of
this paper we use the sub-GeV Kamiokande results.

\subsection{Solar neutrinos}
At present there are four ongoing experiments that are measuring
the flux of solar neutrinos; the Homestake $^{37}{Cl}$ neutrino
capture experiment, the Kamiokande $\nu$-e scattering experiment
and the $^{71}{Ga}$ neutrino capture experiments of the SAGE as well as
the GALLEX collaborations. The measured rate for for the
chlorine experiment \cite{cl} is 2.55 $\pm$ 0.17 (stat) $\pm$ 0.18
(syst) SNU as
compared to the standard model \cite{ssm2} prediction :
9.3$^{+1.3}_{-1.4}$ SNU including metal and helium diffusion and
7.0$^{+0.9}_{-1.0}$ SNU without any diffusion. The observed rates
in the Ga experiments GALLEX \cite{gallex} and SAGE \cite{sage}
are respectively 79 $\pm$ 10 (stat) $\pm$ 6 (syst) SNU and 69 $\pm$ 11
(stat) $^{+5}_{-7}$ (syst) SNU while the
theoretical prediction is 137 $^{+6}_{-7}$ SNU
including diffusion and 126$^{+6}_{-6}$ SNU for no diffusion.
The measured flux in Kamiokande \cite{kam} is [3.0 $\pm$ 0.41 (stat)
$\pm$ 0.35
(syst)] $\times 10^{-6}~ {cm}^{-2}s^{-1}$. This is 0.45 of the
theoretical predictions including diffusion and 0.61 of the standard
model rate without diffusion \cite{ssm2}.
Thus all the experiments indicate deviations from the standard model
predictions.
The degree of depletion differ from experiment to experiment. Since
each type of experiment is sensitive to different parts of the solar
neutrino energy spectrum, it is plausible that the suppression
mechanism is energy dependent. It has remained an unsettled issue for a
long time and new experiments \cite{sno} are being set up to examine it
from various angles.

\section {Neutrino Oscillation}
\subsection{Vacuum Oscillation}
If the neutrinos are massive then quark mixing suggests a
leptonic mixing matrix analogous to the CKM matrix. Then, the flavour
eigenstates, {\it{i.e.}} the states produced in weak interaction
decays are not the same as the mass eigenstates (which are the
neutrino states that propagate) but linear combinations of them.
\begin{equation}
\nu_{\alpha} = \Sigma_{i=1}^{N}{U_{\alpha i} \nu_{i}}
\end{equation}
$U$ is the unitary mixing matrix. This leads to the possibility of
neutrino oscillations.
For N neutrino generations the  expression (\ref{pnu}) generalises
to
\begin{equation}
P_{\nu_{\alpha}\nu_{\beta}} = \delta_{\alpha \beta} - 4~ \Sigma_{j
> i}~ U_{\alpha i} U_{\beta i} U_{\alpha j} U_{\beta j}
\sin^{2}(\frac{\pi L}{\lambda_{ij}})
\label{pab}
\end{equation}
i, j varies from 1 to N for N generations.
$\lambda_{ij} = 2.47m ({E_{\nu}}/{MeV})
({{eV}^{2}}/{\Delta_{ij}}$).
The actual form of the various survival and transition
probabilities will depend on the spectrum of ${\Delta m}^{2}$ chosen
and the explicit form of $U$.

If ${\Delta m}^2$ is such that a particular $\lambda >> L$, then the
corresponding oscillatory term $\sin^2{\pi L/\lambda} \rightarrow 0 $,
whereas $\lambda << L$ would imply a large number of oscillations and
consequently the $\sin^2{\pi L/\lambda}$ term averages out to 1/2.

Since CP violating phases are discarded, $U$ is real and in the minimal
mixing scheme, is a function of four angles. One can express it
in general as the product of four $4 \times 4$ rotation matrices,
$R_{ij}$. The order in which the multiplication is to be performed is
arbitrary leading to several different forms of $U$. In the quark
sector considerable effort has been devoted to specify what would be a
`good choice' for N=3 as well as N $>$ 3 \cite{ckm}. All choices being
mathematically equivalent, a good choice is mostly a matter of
convenience, depending on how directly one can connect the
experimentally determined quantitities with the mixing angles. In the
neutrino sector the experiments measure the energy and detector
cross-section averaged survival or transition probabilities.
Consequently a judicious choice of $U$ in this case would be that in
which one can minimise the number of mixing angles appearing in the
various survival and transition probabilities. With this as the guideline
for determining the mixing matrix $U$, we observe that the order in
which the product of the rotation matrices are to be taken depends on
the $\Delta_{ij}$s chosen. As a general principle we fix the following
rule: the rotations are to be performed in the order of increasing mass
hieararchies. This is true even for a three generation analysis.
As a consequence of CPT and CP invariance there is
no distinction between
$P_{\nu_{\alpha}\nu_{\beta}}, P_{\overline{\nu}_{\alpha}
\overline{\nu}_{\beta}}$, $P_{\nu_{\beta}\nu_{\alpha}},
P_{\overline{\nu}_{\beta}\overline{\nu}_{\alpha}}$ in our analysis.

\subsection{Matter Oscillation}
The discussion on matter oscillations is geared towards the solar
neutrino problem since we neglect the matter effects for the
atmospheric neutrinos. For the mass scales involved in our analysis
this is a good approximation \cite{carlson}.

We have kept $\Delta_{14}$ in the range suitable for solving the solar
neutrino problem for both the mass patterns, implying $\nu_e
- \nu_s$ oscillation as the dominant mode for depletion of solar
neutrinos. In a combined analysis, mixing of $\nu_e$ with $\nu_{\mu}$
and $\nu_{\tau}$ is also expected to affect the probabilities.

Solving the neutrino propagation equations in matter for more than two
generations and arbitrary values of neutrino masses is in general a
non-trivial exercise. MSW analysis for three neutrino generations and the
conditions under which it simplifies have been studied by many authors
\cite{kuo}. A particularly simplifying assumption is one in which
the problem reduces to an effective two generation case
\cite{anjan,fogli1}.
Oscillation to sterile neutrinos as a solution to the solar neutrino
problem has been pursued by many authors in a two generation framework
-- for both vacuum \cite{sterilev}, and matter oscillations
\cite{sterilem}.
In the presence of more than two
generations the propagation equations in matter differ from the two
generation $\nu_e - \nu_s$ case, due to the  charged current
interactions of only $\nu_{e}$ with matter as well as the different
neutral current interactions of the three active flavours and $\nu_{s}$.
In our analysis we follow the notations of \cite{anjan} but
extend it for four generations including the changes in the effective
potentials arising from the presence of sterile neutrinos.

The neutrino propagation equation in matter for four neutrino
generations is,
\begin{equation}
i\frac{d}{dx}{\pmatrix{\nu_{e} \cr \nu_{\mu} \cr \nu_{\tau}
\cr \nu_{x} \cr}} = \frac{1}{2E} M_{F}^{2} {\pmatrix{\nu_{e} \cr \nu_{\mu}
\cr \nu_{\tau} \cr \nu_{x} \cr }}
\end{equation}
where, the mass matrix ${M_{F}}^{2}$ in flavour basis is
\begin{equation}
{M_{F}}^{2} = U {M_{D}}^2 U^{\dagger} + {M_{A}}^2
\label{mf}
\end{equation}
\begin{equation}
{M_{D}}^2 =
{\pmatrix {0 & 0 & 0 & 0 \cr  0 & \Delta_{21} & 0 & 0\cr
0 & 0 & \Delta_{31} & 0 \cr 0 & 0 & 0 & \Delta_{14} \cr}} + {m_{1}}^2
{\bf I}
\end{equation}
\begin{equation}
{M_{A}}^2 = {\pmatrix{ A_{1} & 0 & 0 & 0
\cr 0 & 0 & 0 & 0 \cr 0 & 0 & 0 & 0 \cr 0 & 0 & 0 & A_{2} \cr}}
- A_{2} {\bf I}
\end{equation}
$ A_{1} = {2\sqrt{2}G_{F}n_{e}}E$ and arises from the charged current
interaction of $\nu_{e}$ with electrons while $A_{2} =
{\sqrt{2}G_{F}n_{n}}E$ is due to the neutral current interaction of
$\nu_{e}, \nu_{\mu}$ as well as $\nu_{\tau}$ with the neutrons.
$G_{F}$ is the Fermi coupling constant.
$n_{e}$ and $n_{n}$ are the ambient electron and neutron
density respectively. E is the neutrino energy. For $\Delta_{12}$,
$\Delta_{13} >> A_{1}$, ${M_{F}}^{2}$ can be separated into a 2$\times$2
neutrino submatrix exactly of the two generation form
and two decoupled neutrinos. Under these conditions MSW resonance
occurs between the first and the fourth generation while the
second and the third generation remain unaffected by matter. This
conclusion has been checked in \cite{anjan} for one mass squared
difference in the atmospheric range. Since $\Delta_{LSND} >>
\Delta_{ATM}$, the above arguement is expected to be valid for this
mass scale also. We evaluate the survival probabilities in this
approximation  in the subsequent section for both the mass
patterns.
\section{Survival and Transition Probabilities}
The oscillations are characterised by three oscillation
wavelengths $\lambda_{LSND}$, $\lambda_{ATM}$, $\lambda_{SOLAR}$
corresponding to the three mass scales, $\Delta_{LSND}$,
$\Delta_{ATM}$ and $\Delta_{SOLAR}$ in the problem.
We note the following general points:\\
(i) Accelerator and Reactor Neutrinos\\
For the accelerator and reactor neutrinos, the energy
and length scales are such that $\lambda_{SOLAR}$  and
$\lambda_{ATM} >> L$  (see Table 1) and  the
oscillations driven by these mass scales are absent.
Thus the one mass scale dominance often used in the
context of accelerator and reactor neutrino oscillation
\cite{fogli1,one} is a valid approximation, the oscillations
being driven by $\lambda_{LSND}$. We further note that for Bugey
$\lambda_{LSND} << L$ so that ${\sin^2}(\pi L/\lambda_{LSND})$
averages to 1/2. \\
\noindent{(ii) Atmospheric neutrinos}\\
For the atmospheric neutrinos in the energy range $\sim$ (0.1 -- 1) GeV
travelling through a distance ranging from $\sim$ (10 -- $10^4$) km,
$\lambda_{LSND} << L$ and ${\sin^2}(\pi L/\lambda_{LSND})$
can be replaced by the average value 1/2. On the other hand
$\lambda_{SOLAR} >> L$ and the oscillations driven by the mass scale
relevant for the solar neutrino problem vanish in this case. \\
(iii)Solar neutrinos\\
For oscillation of solar neutrinos in vacuum  $\lambda_{LSND}$ and
$\lambda_{ATM} << L$ and the terms involving these average out to 1/2.\\

\subsection{Mass Spectrum (i)}
Following the prescription described in section 3.1,
\begin{eqnarray}
U  & = & R_{23}R_{13}R_{12}R_{14}
           \nonumber\\
   & = & {\small{\pmatrix {c_{12}c_{13}c_{14} & s_{12}c_{13} &  s_{13} &
          c_{13}c_{12}s_{14} \cr
          -c_{23}s_{12}c_{14} - s_{13}c_{12}s_{23}c_{14} &
          c_{12}c_{23} - s_{13}s_{12}s_{23} & s_{23}c_{13} &
           -c_{23}s_{12}s_{14} - s_{13}c_{12}s_{23}s_{14} \cr
           s_{12}s_{23}c_{14} - s_{13}c_{12}c_{23}c_{14} &
           -c_{12}s_{23} - s_{13}s_{12}c_{23} & c_{13}c_{23} &
           s_{12}s_{23}s_{14} - s_{13}c_{12}c_{23}s_{14} \cr
           -s_{14} & 0 & 0 & c_{14} \cr}}}
            \nonumber\\
\label{u2}
\end{eqnarray}
where $c_{ij} =\cos{{\theta}_{ij}}$ and $s_{ij} =\sin{{\theta}_{ij}}$,
here and everywhere else in the paper.
Let us now see what the probabilities for the various experiments are
in this case. \\
\noindent{(i) Accelerator and Reactor Experiments}
\begin{equation}
P_{\overline{\nu}_e\overline{\nu}_e} = 1 - 2s_{13}^2c_{13}^2~~~(Bugey)
\label{b2}
\end{equation}
\begin{equation}
P_{\overline{\nu}_{\mu}\overline{\nu}_{\mu}} = 1 - 4s_{23}^2c_{13}^2(1
- c_{13}^2s_{23}^2){\sin^2}({\pi L/\lambda_{LSND}})~~~(CDHSW)
\label{cdhsw2}
\end{equation}
\begin{equation}
P_{\overline{\nu}_{\mu}\overline{\nu}_e} = 4 c_{13}^2 s_{13}^2 s_{23}^2
{\sin^2}({\pi L/\lambda_{LSND}})~~~(LSND, E776)
\label{lsnd2}
\end{equation}
\begin{equation}
P_{\overline{\nu}_{\mu}\overline{\nu}_{\tau}} = 4 c_{23}^2 s_{23}^2 c_{13}^4
{\sin^2}({\pi L/\lambda_{LSND}})~~~(E531, CHORUS, NOMAD)
\label{chorus1}
\end{equation}
\noindent{(ii) Atmospheric neutrinos}\\
The general expression of (\ref{ratm}) for N flavours in terms of
the neutrino transition and survival probabilities is
\begin{equation}
R = { \frac{ P_{\nu_\mu \nu_\mu} + r_{MC} P_{\nu_\mu
\nu_e}}{P_{\nu_e \nu_e} + \frac{1}{r_{MC}} P_{\nu_\mu \nu_e}}}
\label{r3}
\end{equation}
where $r_{MC}$ = $(\nu_e + \overline{\nu}_e)/(\nu_{\mu} +
\overline{\nu}_{\mu})$ as obtained from a Monte-Carlo simulation.
In this case the relevant probabilities appearing in (\ref{r3}) are
\begin{equation}
P_{\nu_e \nu_e} = 1 - 2c_{13}^{2}s_{13}^2  - 4c_{12}^2s_{12}^2c_{13}^4
{\sin^2}({\pi L/\lambda_{ATM}})
\label{a1}
\end{equation}
\begin{equation}
P_{\nu_\mu \nu_e} = 2c_{13}^2s_{13}^2s_{23}^2 + 4c_{13}^2c_{12}s_{12}
(c_{23}c_{12} - s_{13}s_{12}s_{23})(c_{23}s_{12} + s_{13}c_{12}s_{23})
{\sin^2}({\pi L/\lambda_{ATM}})
\label{a2}
\end{equation}
\begin{equation}
P_{\nu_\mu \nu_\mu} = 1 - 2c_{13}^{2}s_{23}^2 + 2{c_{13}^4}{s_{23}^4}
 - 4{\sin^2}({\pi L/\lambda_{ATM}})
{(c_{23}c_{12} - s_{13}s_{12}s_{23})^2}{(c_{23}s_{12} + s_{13}c_{12}s_{23})^2}
\label{a3}
\end{equation}
\noindent{(iii)Solar neutrinos}\\ \\
\noindent{(a) Vacuum Oscillations}\\
The electron neutrino survival probability in this case is,
\begin{equation}
P_{\nu_e \nu_e} = c_{13}^4c_{12}^4 P_{2VAC} + s_{13}^4 + s_{12}^4
c_{13}^4
\label{pvac2}
\end{equation}
where $P_{2VAC}$ is of the form of the two generation vacuum
oscillation  probability:
\begin{equation}
P_{2VAC} = 1 - sin^2{2\theta_{14}}\sin^2(\pi L/\lambda_{SOLAR})
\label{p2VAC}
\end{equation}
\noindent{(b) MSW oscillations}\\
If the mass hieararchies, mixing angles and the density distributions
are such that one has resonance between the first and the fourth mass
eigenstates, while the second and the third mass eigenstates remain
independent of matter density then the mixing matrix in matter is
$U_{M} = R_{23}R_{13}R_{12}R_{14M}$.
The mixing between the first and the fourth generation gets modified by
the matter effects as,
\begin{equation}
tan 2\theta_{14m} =  \frac{\Delta_{14}\sin 2\theta_{14}}{\Delta_{14}\cos
2\theta_{14} - 2\sqrt{2}G_{F}n_{eff}E}
\label{14m}
\end{equation}
where we define $n_{eff}$ as
\begin{equation}
n_{eff} = c_{13}^2c_{12}^2n_{e} - {\frac{1}{2}}n_{n}
\label{neff}
\end{equation}
In the limit $c_{13}$, $c_{12}$ $\rightarrow$ 1 this reduces
to the two generation limit for oscillation between $\nu_{e}$ and
$\nu_{s}$ \cite{sterilem}.
{}From (\ref{14m}), the resonance condition between the
first and the fourth generation in the presence of the other two
generations becomes
\begin{equation}
2\sqrt{2}G_{F}n_{eff}E= \Delta_{14}cos{2\theta}_{14}
\label{reso}
\end{equation}
The difference of this with the three generation resonance is to be
noted. While the three generation resonance condition gets modified by
one additional mixing angle here the mixing angles with the second as
well as the third generation appear in the resonance condition through
$n_{eff}$.
The calculation of the survival
probability  $P_{\nu_e \nu_e}$ is then a straight-forward
generalisation of the standard two or three generation MSW scenario
and one gets,
\begin{equation}
P_{\nu_e \nu_e} =c_{12}^4 c_{13}^4 P_{MSW} + s_{13}^4 + s_{12}^4 c_{13}^4
\label{pmsw2}
\end{equation}
\begin{equation}
P_{MSW} = 0.5 + (0.5 - \theta(E-E_{A})X)cos 2\theta_{14} cos 2\theta_{14M}
\label{p2}
\end{equation}
$E_{A}$ = ${{\Delta_{14}}cos2\theta_{14}}/{2\sqrt{2}G_{F}n_{eff}}$
gives the minimum $\nu$ energy that can encounter a resonance
inside the sun.
For this case the jump probability $X$, between the first and the fourth
mass eigenstates, as obtained in the Landau-Zener approximation is
\begin{equation}
X = exp~(-{\pi\Delta_{14}}{\sin^{2}2\theta_{14}}/
{4E{cos2\theta_{14}}\frac{1}{n_{eff}}{(\frac{dn_{eff}}{dx})_{res}}})
\label{x}
\end{equation}\\
We note that unlike the case discussed in \cite{anjan}, the mixing
angles $c_{12}^2$, $c_{13}^2$ appear in the expression of the jump
probability, via $n_{eff}$ defined in (\ref{neff}).

\subsection{Mass Spectrum (ii)}
This case has been considered also in \cite{garcia}, following
slightly different notations. We discuss this to make the paper
self contained and to facilitate the comparison between the results
obtained for the two mass patterns.
Following the rule we specified $U$ in this case is
\begin{eqnarray}
U  & = & R_{13}R_{12}R_{23}R_{14}
           \nonumber\\
   & = & {\pmatrix {c_{12}c_{13}c_{14} & s_{12}c_{13}c_{23} - s_{13}s_{23}
         & c_{13}s_{12}s_{23} + s_{13}c_{23} & c_{13}c_{12}s_{14} \cr
         -s_{12}c_{14} & c_{12}c_{23} &c_{12}s_{23} & -s_{12}s_{14} \cr
          -s_{13}c_{12}s_{14} & -s_{13}s_{12}c_{23} - c_{13}s_{23} &
           -s_{12}s_{13}s_{23} + c_{13}c_{23} & -s_{13}c_{12}s_{14} \cr
            -s_{14} & 0 & 0 & c_{14} \cr}}
\label{u1}
\end{eqnarray}
This reduces to the three generation case discussed in \cite{ska} in
the limit of $\theta_{14}$ going to zero. We next calculate the
probabilities explicitly for the various experiments.\\
\noindent{(i) Accelerator and Reactor Experiments}\\
\begin{equation}
P_{\overline{\nu}_e\overline{\nu}_e} = 1 - 2c_{13}^2c_{12}^2 +
2c_{13}^4c_{12}^4~~~(Bugey)
\label{bugey}
\end{equation}
\begin{equation}
P_{\overline{\nu}_{\mu}\overline{\nu}_{\mu}} = 1 - \sin^2{2\theta_{12}}
{\sin^2}({\pi L/\lambda_{LSND}})~~~(CDHSW)
\end{equation}
\begin{equation}
P_{\overline{\nu}_{\mu}\overline{\nu}_e} = 4 c_{12}^2 s_{12}^2 c_{13}^2
{\sin^2}({\pi L/\lambda_{LSND}})~~~(LSND, E776 )
\label{pnumul}
\end{equation}
\begin{equation}
P_{\overline{\nu}_{\mu}\overline{\nu}_{\tau}} =
4 c_{12}^2 s_{12}^2 s_{13}^2
{\sin^2}({\pi L/\lambda_{LSND}})~~~(E531, CHORUS, NOMAD)
\label{chorus2}
\end{equation}
\noindent{(ii) Atmospheric neutrinos}\\
For the chosen mass pattern and mixing the  probabilities appearing
in (\ref{r3}) are
\begin{equation}
P_{\nu_e \nu_e} = 1 - 2c_{13}^{2}c_{12}^2 + 2c_{13}^{4}c_{12}^4 - 4
(c_{13}s_{12}c_{23} - s_{13}s_{23})^{2} (c_{13}s_{12}s_{23} +
s_{13}c_{23})^{2}{\sin^2}({\pi L/\lambda_{ATM}})
\label{pnueatm}
\end{equation}
\begin{equation}
P_{\nu_\mu \nu_e} = 2c_{13}^2c_{12}^2s_{12}^2 - 4c_{12}^2c_{23}s_{23}
(c_{13}s_{12}c_{23} - s_{13}s_{23})(c_{13}s_{12}s_{23} + s_{13}c_{23})
{\sin^2}({\pi L/\lambda_{ATM}})
\label{pmueatm}
\end{equation}
\begin{equation}
P_{\nu_\mu \nu_\mu} = 1 - 2c_{12}^{2}s_{12}^2 -
4c_{12}^4c_{23}^2s_{23}^2{\sin^2}({\pi L/\lambda_{ATM}})
\label{pnumuatm}
\end{equation}

\noindent{(iii)Solar neutrinos}\\ \\
\noindent{(a) Vacuum Oscillations}\\
The electron neutrino survival probability in this case is,
\begin{equation}
P_{\nu_e \nu_e} = c_{13}^4c_{12}^4 P_{2VAC} +
(s_{12}c_{13}c_{23} - s_{13}s_{23})^4 + (c_{13}s_{12}s_{23} +
s_{13}c_{23})^4
\label{pvac1}
\end{equation}
where $P_{2VAC}$ is given by eqn. (\ref{p2VAC}).\\
\noindent{(b) MSW oscillations}\\
In this case the mixing matrix in matter is,
$U_{M} = R_{13}R_{12}R_{23}R_{14M}$.
For this also the resonance can be assumed to happen between the first
and the fourth mass eigenstates whence the mixing angle
$\theta_{14M}$ and the resonance condition continue to be given by eqns.
(\ref{14m}) and (\ref{reso}). The probability however is different and
can be expressed as,
\begin{equation}
P_{\nu_e \nu_e} =c_{12}^4 c_{13}^4 P_{MSW} + (s_{12}c_{13}c_{23} -
s_{13}s_{23})^4 + (c_{13}s_{12}s_{23} + s_{13}c_{23})^4
\label{pmsw1}
\end{equation}
where $P_{MSW}$ is defined in eqn. (\ref{p2}).\\

\section{Results and Discussions}
For both the mass spectrums discussed in sections 4.1 and 4.2, the
survival or transition
probabilities for the accelerator and reactor
experiments are functions of any two of the mixing angles --  $\theta_{12},
\theta_{13}, \theta_{23}$ --
and one mass squared difference, $\Delta_{LSND}$. In a realistic
analysis one has to average the
probabilities over the $L/E$ distributions of various experiments and
fold it with the detector cross-sections.
We adopt the approach followed in \cite{fogli1} since one mass scale
dominance is a good approximation in our case also. As noted in
\cite{fogli1} in this limit one can make a one to one
correspondence between $\sin^2{2\theta}$ as obtained from a two flavour
analysis and the angular factor appearing in a three generation
calculation. If we now fix $\Delta{m}^2$ around $\Delta_{LSND} \sim$ 6
$eV^2$ then from the
bound on  $\sin^2{2\theta}$ at this $\Delta{m}^2$ from two
flavour exclusion contours one can constrain the three generation mixings.
While the other experiments present the
exclusion contours at 90\% C.L., LSND gives their plots at 95\% C.L.
\cite{lsnd}. Thus for this we use the quoted value for the probability at
90\% C.L. and use the limit ${\sin^2}({\pi L/\lambda_{LSND}})
\rightarrow 1$.

The probabilities in the atmospheric neutrino case are functions of
the three mixing angles  $\theta_{12},
\theta_{13}, \theta_{23}$ and one mass squared difference.
In our analysis we approximate the
$\sin^2({\pi L/\lambda_{ATM}})$ factor by its averaged value 0.5
as is often done in the context of the sub-GeV
data \cite{barger,pakvasa,acker}.
This can be improved by an averaging over the
incident neutrino energy  spectrum, the zenith-angle of the beam as
well as the final lepton energy \cite{barger,fogli1}. $r_{MC}$ is
taken to be 0.45 from a detailed Monte-Carlo simulation including
the effects of muon polarisation \cite{gaisser}.
Using the sub-GeV Kamiokande results:
\begin{equation}
0.48 \leq R \leq 0.73 ~(90\% {\rm C.L.})
\label{r90}
\end{equation}
We determine how much of the area, admissible from neutrino oscillation
searches at laboratory, permits $R$ to lie in the above range for fixed
values of the third mixing angle. The allowed ranges of
this mixing angle consistent with the atmospheric data are also
obtained.

Finally we check the compatibility of the mixing angles found from this
combined analysis with the solar neutrino results. The sensitivity of
CHORUS and NOMAD in these areas are explored.
Below we discuss the results for each mass spectrum.\\

\noindent{Mass Spectrum (i)}\\
The various constraints in this case from the accelerator and reactor
results are:
\begin{equation}
2 s_{13}^2 c_{13}^2 < 0.05~~~ (Bugey)
\label{bm2}
\end{equation}
\begin{equation}
4s_{23}^2c_{13}^2(1 - c_{13}^2s_{23}^2)  <  0.08~~~(CDHSW)
\end{equation}
\begin{equation}
4 s_{13}^2 c_{13}^2 s_{23}^2 <  0.003~~~(E776)
\end{equation}
\begin{equation}
0.0002 \leq 4 s_{13}^2 c_{13}^2s_{23}^2  \leq  0.0069 ~~~(LSND)
\end{equation}
\begin{equation}
4 s_{23}^2 c_{23}^2 c_{13}^4 <  0.02~~~(E531)
\end{equation}
{}From the Bugey constraint (\ref{bm2})
\begin{center}
$s_{13}^2  \stackrel{\textstyle{<}}{\sim} $ 0.026 or
$s_{13}^2$ $\stackrel{\textstyle{>}}{\sim}$ 0.974.
\end{center}
Confinining $s_{13}^2$ within these limits we scan the whole range
$0 \leq s_{23}^2 \leq 1$ to determine the admitted area in the $s_{13}^2 -
s_{23}^2$ plane consistent with the accelerator and reactor experiments
listed above. A large portion of the parameter space can be
ruled out combining all the restrictions and three allowed sectors are
obtained -- presented in figs. 2a, 2b and 2c respectively.

Fig. 2a shows the zone where  0.974 $ \stackrel{\textstyle{<}}{\sim}
s_{13}^2 <$ 1.0. As is clear from the figure, the most stringent limit
on $s_{23}^2$, in this range, comes from E776 and LSND constraints.
The permitted area of fig. 2a corresponds to the case
$s_{13}^2 \rightarrow $ 1.
In this regime the relevant probabilities for atmospheric neutrinos
given by eqns. (\ref{a1}) to (\ref{a3}) can be expressed as $P_{\nu_e
\nu_{e}} \simeq $ 1, $P_{\nu_e \nu_{\mu}} \simeq $ 0, $P_{\nu_{\mu}
\nu_{\mu}} \simeq  1 - \frac{1}{2}\sin^2{(2\theta_{12} +
2\theta_{23})}$.  This is the  $\nu_{\mu} - \nu_{\tau}$ oscillation
limit, where consistent solutions to the atmospheric puzzle may be
obtained. However for the solar neutrino survival probability, eqn.
(\ref{pmsw2}), the limit $s_{13}^2 \rightarrow $ 1 would imply that the
coefficient of the vital term responsible for the MSW effect,
$P_{MSW}$, becomes very small. Consequently $P_{\nu_e \nu_e}
\rightarrow$ 1 due to the factor $s_{13}^4$ in eqn. (\ref{pmsw2})
contrary to the results from the solar neutrino experiments discussed
in section 2.3.
Similarly for the vaccuum oscillation probabilitiy (eqn. (\ref{pvac2}))
also one would require that the factor $c_{13}^4 c_{12}^4$ multiplying
the energy and ${\Delta m}^2$ dependent term, $P_{2VAC}$ should not be
too small which would again prefer low $s_{13}^2$.
Thus this area is disfavoured by the solar neutrino data.

For low $s_{13}^2$ ($\stackrel{\textstyle{<}}{\sim}$ 0.026) the
allowed values of $s_{23}^2$ is severely constrained by E531 result
and only very high ($>$ 0.99) or very low ($<$ 0.01)  $s_{23}^2$ are
admissible.

Fig. 2b shows the low $s_{13}^2$ -- high $s_{23}^2$ zone. In this
regime E776 data puts a stronger constraint on $s_{13}^2$ than Bugey
and sets the limit  $s_{13}^2 <$ 0.001.
In this part of the parameter space a consistent solution to the
atmospheric neutrino anomaly cannot be found. This region corresponds
to $s_{13}^2 \rightarrow$ 0, $s_{23}^2 \rightarrow$ 1, whence from
eqns. (\ref{a1}) to (\ref{a3})
$P_{\nu_e \nu_{e}} \simeq 1 - 2c_{12}^2s_{12}^2$, $P_{\nu_e \nu_{\mu}}
\simeq $ 0 and $P_{\nu_{\mu} \nu_{\mu}} \simeq $ 1. Thus in this zone
$\nu_{e} - \nu_{\tau}$ oscillations take place
driving the ratio of ratios $R$ in a direction opposite to that required.

Fig. 2c is the  low $s_{13}^2$ -- low $s_{23}^2$ region.
In this portion of the parameter space most severe restrictions are
from E531, LSND and Bugey  giving the following constraints:
0.01 $< s_{13}^2 < $ 0.026, 0.002 $< s_{23}^2 <$ 0.005.
The angles in this  sector can be approximated as
$s_{13} \rightarrow $ 0, $s_{23} \rightarrow 0$. In this limit the
relevant probabilities for the atmospheric neutrinos (eqns. (\ref{a1})
to (\ref{a3}))  are  $P_{\nu_e \nu_{e}} \simeq 1 - 2c_{12}^2s_{12}^2$,
$P_{\nu_e \nu_{\mu}} \simeq  2c_{12}^2s_{12}^2$ and
$P_{\nu_{\mu}\nu_{\mu}} \simeq  1 - 2c_{12}^2s_{12}^2$. Thus this
belongs approximately to the $\nu_e - \nu_{\mu}$ oscillation region
where the atmospheric anomaly can be explained consistently.
Varying $s_{13}^2$ and $s_{23}^2$ in the range shown in fig. 2c the
limits on  $s_{12}^2$ compatible with the atmospheric neutrino
constraint (\ref{r90}) are  presented in fig. 3. Two
allowed bands are obtained for $s_{12}^2$:
0.10 $\stackrel{\textstyle{<}}{\sim} s_{12}^2
\stackrel{\textstyle{<}}{\sim}$ 0.37 and
0.61 $\stackrel{\textstyle{<}}{\sim} s_{12}^2
\stackrel{\textstyle{<}}{\sim}$ 0.88. Taking six different
$s_{12}^2$s, from  the above range corresponding to the edges of the two
permitted zones and the parts inside them, the allowed area in the
$s_{13}^2 - s_{23}^2$ plane from accelerator, reactor and atmospheric
neutrino data is shown shaded in fig. 4.
Compatibility with the solar neutrino data disfavours the high
values of $s_{12}^2$ and only part of the parameter spaces shown in
figs. 2c, 3 and 4 satisfy all the experimental constraints.

Fixing  $\Delta_{LSND} \sim$ 6 eV$^2$, from eqn. (\ref{chorus1})
$P_{\nu_{\mu} \nu_{\tau}} = 0.16s_{23}^2c_{23}^2c_{13}^4$ for CHORUS
and NOMAD.
For typical values of mixing angles from this combined allowed zone,
namely, $s_{13}^2$ = 0.003 and $s_{23}^2$ = 0.02,
$P_{\nu_{\mu}\nu_{\tau}} \sim $ 3.12 $\times$ 10$^{-3}$, which is
greater than the minimum sensitivity, $10^{-4}$, of these experiments
and can be marginally within their reach.\\

\noindent{Mass Spectrum (ii)} \\
In this case the constraints on mixing angles at $\Delta_{LSND}$ $\sim$
6 $eV^2$ from the various laboratory experiments are:
\begin{equation}
2c_{13}^2c_{12}^2 - 2c_{13}^4c_{12}^4 < 0.05~~~ (Bugey)
\label{c1}
\end{equation}
\begin{equation}
4c_{12}^2s_{12}^2  < 0.08~~~(CDHSW)
\label{c2}
\end{equation}
\begin{equation}
4 c_{12}^2 s_{12}^2 c_{13}^2 < 0.003~~~(E776)
\label{c3}
\end{equation}
\begin{equation}
0.0002 \leq 4 c_{12}^2 s_{12}^2 c_{13}^2 \leq 0.0069 ~~~(LSND)
\label{c4}
\end{equation}
\begin{equation}
4 c_{12}^2 s_{12}^2 s_{13}^2 < 0.02~~~(E531,~~ \rm{from}~~ \nu_{\mu}
\rightarrow \nu_{\tau})
\end{equation}
For this case also three allowed areas in the relevant
$s_{12}^2 -  s_{13}^2$ plane are obtained, from the
laboratory constraints. The experiments
which are most restrictive in this case can be different in general from
the mass spectrum (i). These regions are displayed in figs. 5a, b and c
respectively.
The CDHSW constraint (\ref{c2}) gives,
\begin{center}
$s_{12}^2  \stackrel{\textstyle{<}}{\sim} $ 0.02 or
$s_{12}^2$ $\stackrel{\textstyle{>}}{\sim}$ 0.98.
\end{center}

Fig. 5a shows the area for which $s_{12}^2$
$\stackrel{\textstyle{>}}{\sim}$ 0.98. Allowed ranges of $s_{13}^2$ in
this region are determined by E776 and LSND.
This corresponds to $s_{12}^2 \rightarrow$ 1. In this
limit for the atmospheric neutrinos
$P_{\nu_e \nu_{\mu}} \simeq$ 0  and
$P_{\nu_{\mu}\nu_{\mu}} \simeq $ 1, from the expressions
(\ref{pmueatm}) and (\ref{pnumuatm}). Thus this is the $\nu_e -
\nu_{\tau}$ oscillation regime and is not consistent with the
atmospheric anomaly.

For $s_{12}^2
\stackrel{\textstyle{<}}{\sim} $ 0.02, the Bugey constraint (\ref{c1})
restricts the permissible values of $s_{13}^2$ to
be $\stackrel{\textstyle{>}}{\sim}$ 0.97 or
$\stackrel{\textstyle{<}}{\sim}$ 0.03.

Fig. 5b contains the area where $s_{12}^2
\stackrel{\textstyle{<}}{\sim} $ 0.02 and
$s_{13}^2$ $\stackrel{\textstyle{>}}{\sim}$ 0.97.
In this region $s_{13}^2 \rightarrow$ 1, $s_{12}^2
\rightarrow $ 0. Then, the probabilities given by eqns. (\ref{pnueatm})
to (\ref{pnumuatm}) for the atmospheric neutrinos assume the following forms:
$P_{\nu_e \nu_{e}} \simeq 1 - 2c_{23}^2s_{23}^2$, $P_{\nu_e \nu_{\mu}}
\simeq 2 c_{23}^2 s_{23}^2$  and $P_{\nu_{\mu} \nu_{\mu}} \simeq 1
- 2 c_{23}^2 s_{23}^2$. Thus, in this limit $P_{\nu_{\mu}\nu_{\tau}}
\simeq 0$ and the atmospheric puzzle can be explained by $\nu_e -
\nu_{\mu}$ transitions. However, since $s_{13}^2 \rightarrow$ 1,
this region is incompatible with the solar neutrino flux measurements
as is evident from  (\ref{pvac1}) and (\ref{pmsw1}).

Fig. 5c shows the area 0 $< s_{12}^2 < 8 \times 10^{-4}$ and 0 $<
s_{13}^2 <$ 0.03. Here, E776 data puts a tighter bound on $s_{12}^2$
than CDHSW. The constraint on $s_{13}^2$ is determined by Bugey.
This region corresponds to the limit $s_{13}^2 \rightarrow$ 0,
$s_{12}^2 \rightarrow $ 0. In this range eqns. (\ref{pnueatm})
to (\ref{pnumuatm}) for atmospheric neutrinos imply
$P_{\nu_e \nu_{e}} \simeq 1 $,
$P_{\nu_e \nu_{\mu}} \simeq $ 0 and $P_{\nu_{\mu} \nu_{\mu}} \simeq$ 1
- 2 $c_{23}^2 s_{23}^2$. So this corresponds to
$\nu_{\mu} - \nu_{\tau}$ oscillations in the atmosphere.
Substituting these in (\ref{r90}) one gets the limits on $s_{23}^2$ as
0.162 $< s_{23}^2 < $ 0.838. A similar situation was discussed in \cite{ska},
which considered the three generation limit of the mixing matrix (\ref{u1}).
As in \cite{ska} for all $s_{23}^2$ lying within this limit the whole
of the parameter space shown in fig. 5c is consistent with the
condition (\ref{r90}).
Thus in fig. 6 we present the allowed region in the $s_{12}^2$ --
$s_{13}^2$ plane consistent with accelerator, reactor and atmospheric
neutrino data choosing one representetive value of $s_{23}^2$ (= 0.2)
from the above range.
The difference with \cite{ska} is, here, E776 constraints further
narrows down the allowed range of $s_{12}^2$. Since in this zone both
$s_{12}^2$ and $s_{13}^2$ stay close to zero this is consistent with
the solar neutrino results.

For CHORUS and NOMAD from eqn. (\ref{chorus2}) $P_{\nu_{\mu}\nu_{\tau}} =
0.16s_{12}^2c_{12}^2s_{13}^2$. Taking two typical values of
$s_{12}^2$ and $s_{13}^2$ , namely, $s_{12}^2$ = $10^{-4}$ and
$s_{13}^2$ = 0.02, from the combined allowed zone,
$P_{\nu_{\mu}\nu_{\tau}} \sim 10^{-6}$ which is below the minimum
sensitivity attainable in these experiments.

We observe that in both cases, the solar neutrino survival probailities
eqns.(\ref{pvac2}) and (\ref{pvac1}) for the vacuum oscillation case and
eqns. (\ref{pmsw2}) and (\ref{pmsw1}) for matter oscillations depend on
$\Delta_{14}, \theta_{14}$ as well as on combinations of
$\theta_{12}, \theta_{13}, \theta_{23}$ which depend on the mass pattern.
Fixing the values of these mixing angles in the region determined
by the atmospheric and laboratory results one can find the allowed area
in the $\Delta_{14} - \sin^2{2\theta_{14}}$ plane using the solar
neutrino data. The solar neutrino probabilities in this case differ
from the two generation $\nu_e - \nu_s$ oscillations case, due to the
presence of the mixing angles with the other generations. Also, in the
two generation case involving just $\nu_e$ and $\nu_s$ Kamiokande would
be sensitive to $\nu_e$s only but here one has the additional
possibility of a simultaneous transition to $\nu_{\mu}$s as well as
$\nu_{\tau}$s which can interact in the Kamiokande detector by virtue
of their neutral current interactions.
It has been shown in \cite{anjan,fogli1} that the appearance of
one mixing angle in the expression of probability leads to a larger
area in the MSW  parameter space. Similar conclusions might be obtained
here also. The following points of differences
are to be noted:\\
(i) In \cite{anjan} or \cite{fogli1} three generations are involved
and the resonance condition as well as the survival probabilities are
affected by the presence of only one additional mixing angle, whereas
here one has two or three such mixings.\\
(ii) In the above references oscillation between
active species were considered and hence the term involving the neutron
density was absent and the jump-probability retained its two generation
form. Here due to the asymmetric interaction between the active and the
sterile species the jump-probability between the first and the fourth
state is also affected by mixing with the other generations.

A definitive prediction regarding how these would change the two
flavour allowed zones need a detailed numerical
analysis of the solar neutrino data including the neutrino fluxes,
density profile, interaction cross-sections and a thorough treatment of
the theory errors and their correlations. This is not performed here.

Similarly in the vacuum oscillation case also one can probe whether the
presence of the other mixing angles will alter the two flavour
parameter space.

We note that for the mass spectrum (ii) in the combined allowed zone
depicted in fig. 6, both $s_{12}^2$ and $s_{13}^2$
stay close to 0. Thus in this case, from eqns. (\ref{pvac1}) and
(\ref{pmsw1})  for the vacuum and matter oscillation case respectively,
the presence of the other mixing angles is not expected to change the
two flavour parameter space significantly.

\section{Summary and Conclusions}
We have performed a combined analysis of the accelerator, reactor,
atmospheric and solar neutrino data in a four generation framework
introducing a sterile neutrino, $\nu_{s}$. In such a scenario there are
in general six mass squared differences, three of which
are independent and six mixing angles, neglecting CP violation in the
lepton sector. We assume that $\nu_s$ mixes only with $\nu_e$, thus
reducing the number of mixing angles to four -- $\theta_{12},
\theta_{13}, \theta_{23}$ and $\theta_{14}$. Fixing the three
independent ${\Delta m}^2$s around the ranges from two generation
analyses of the LSND, atmospheric and solar neutrino data, we determine
the mixing angles consistent with all the experimental constraints. We
consider a picture where $\Delta_{14}$ is fixed in the solar neutrino
range (either MSW or vacuum oscillation). Then one can think of two
different mass patterns for the remaining five ${\Delta m}^2$s -- the
mass spectrum (i) in which two ${\Delta m}^2$s are in the atmospheric
range and the other three in the LSND range and the mass spectrum (ii)
where one ${\Delta m}^2$ is in the atmospheric range and the remaining
four in the LSND range. For both cases one can parametrise the mixing
matrix in such a way that the probabilities for the accelerator and
reactor experiments are functions of only two mixing angles and one
independent ${\Delta m}^2$ viz $\Delta_{LSND}$. Fixing $\Delta_{LSND}$
at 6$eV^2$ we map out the allowed zone in the $\sin^2{\theta_{13}} -
\sin^2{\theta_{23}}$ ($\sin^2{\theta_{12}} - \sin^2{\theta_{13}}$)
plane for the mass spectrum (i) ((ii)). Using the atmospheric neutrino
constraint the above area can be further restricted and the permissible
ranges for the remaining mixing angle can be determined. Next we
examine whether the combined allowed area thus obtained is compatible
with the solar neutrino results.

In general for both mass patterns the following picture emerges:
the accelerator and reactor
experiments give three allowed sectors of relevant mixing
angles. In two of these zones a simultaneous solution to the
atmospheric anomaly is possible. One among these is disfavoured by the
solar neutrino data -- leaving us with a narrow range for permitted
$\theta_{12}, \theta_{13}$ and $\theta_{23}$.
For the mass spectrum (i) the admitted zone from all the input
information is the one where the atmospheric puzzle can be explained
by $\nu_e - \nu_{\mu}$ oscillation, while for the mass spectrum (ii) in
the combined allowed zone it is due to $\nu_{\mu} - \nu_{\tau}$ oscillations.
For both mass patterns $P_{\nu_{\mu}\nu_{\tau}}$
for CHORUS and NOMAD is much below the minimum reach of these
experiments in the region where the solution to the atmospheric problem
is via $\nu_{\mu} - \nu_{\tau}$ oscillations. For mass spectrum (ii)
this being the combined allowed zone, cannot be explored by CHORUS and NOMAD
whereas for (i) in the favoured zone
$P_{\nu_{\mu}\nu_{\tau}}$ for CHORUS and NOMAD can be greater than the
minimum sensitivity of $10^{-4}$  and could be probed by these
experiments.

In conclusion we would like to mention that though a three flavour
mixing scheme cannot accommodate the three hieararchically different
mass ranges required for LSND, atmospheric and solar neutrino
oscillations, in a four generation framework with an
additional sterile neutrino there are more than one possible mass
spectrums that can account for all the data simultaneously. We
discussed two such mass spectrums. In both cases our analysis assumes
that the solar neutrino oscillation is driven mainly by $\nu_e - \nu_s$
transitions. If this scenario is confirmed by the future solar neutrino
experiments and the other experimental inputs do not change
significantly as more data accumulates then it is necessary to go to a
four generation picture. We have shown  that the implications of CHORUS
and NOMAD are different in the two cases and thus they can distinguish
between  these.

In this article the allowed areas are obtained by fixing $\Delta_{LSND}
\sim $ 6 $eV^2$. LSND is sensitive to the range  1-10 $eV^2$ and it
remains to be seen what best-fit value, consistent with KARMEN and
BNL-E776, emerges when more data is accrued. We beleive that the
general conclusions obtained in this analysis will remain the same
for any other value of ${\Delta m}^2$ in the above range, though the
precise values of the allowed mixing angles may be different.

\vskip 30pt

\parindent 0pt

The author is indebted to Dr.Amitava Raychaudhuri for many useful
suggestions, discussions, a careful scrutiny of the manuscript and
encouragement at every stage of this work. She also wishes to thank Dr.
Kamales Kar for discussions, help and encouragement. Financial support
from the  Council of Scientific and Industrial Research, India is
acknowledged.

\newpage
\begin{description}
\item{Table 1:} The characteristics of the most restrictive accelerator
and reactor experiments. $\lambda_{LSND}$ and $\lambda_{ATM}$ are
calculated for $\Delta_{LSND} \sim $ 6 $eV^2$ and $\Delta_{ATM} \sim$
10$^{-2}$ $eV^2$ respectively.
\end{description}
\[
\begin{array}{|c|c|c|c|c|} \hline
{\rm Experiment} & {E} & {L} & {\lambda_{LSND}} & {\lambda_{ATM}}
 \\ \hline
{\rm Bugey} & \sim 5~MeV & \sim 40~m & \sim 2.08~m & 1250~m \\ \hline
{\rm CDHSW} & 2 < E < 20~ GeV & \sim 1~km & (0.83 - 8.33)~ km & (500 -
5000)~km  \\ \hline
{\rm E776} & 1 - 10 ~GeV & \sim 1~km & (0.416 - 4.16)~km & (250 -
2500)~km \\ \hline
{\rm E531} & \sim 50~GeV & 0.949~ km & \sim 22~km & \sim 12500~km
\\ \hline
{\rm LSND} & (36 - 60)~MeV & 30~m & (15 - 25)~m & (9 -15) km \\ \hline
{\rm CHORUS/NOMAD} &  30~GeV & 0.8~km & 12.5~km &
7500~km \\ \hline
\end{array}
\]

\newpage
\begin{center} {\Large{\bf FIGURE CAPTIONS}}
\end{center}
\vskip 30pt
Figure 1: The level diagrams showing the possible mass hieararchies (not
to scale).\\
Figure 2: The allowed region in the $s_{13}^2 - s_{23}^2$ plane from
accelerator and reactor data for the mass pattern (i)\\
(a) The region between the solid lines is allowed by LSND; the area to
the right of the big-dashed line is permitted by E776 and that below the
small-dashed line is allowed from CDHSW. \\
(b) The admiited area from LSND is between the solid lines while that
from CDHSW and E-531 is above the small-dashed and medium-dashed lines
respectively. The area to the left of the big-dashed line is allowed
from E776.  \\
(c) The area between the solid lines is allowed by LSND while that
below the big-dashed, small-dashed and medium-dashed lines are allowed
from E776, CDHSW and E531 respectively.\\
In each of these the combined allowed area is marked `allowed'.\\
Figure. 3: The allowed region in the $s_{12}^2 - s_{23}^2$ plane
consistent with accelerator, reactor and atmospheric neutrino
constraints. $s_{13}^2$ is varied in the range determined from fig.
2(c). \\
Figure 4: The allowed area of fig. 2c that is consistent with the
atmospheric neutrino constraint is shown shaded for six different
values of $s_{12}^2$ from the admiited range in fig. 3.\\
Figure 5: The allowed region in the $s_{12}^2 - s_{13}^2$ plane from
accelerator and reactor data for the mass pattern (ii)\\
(a) The area between the curved solid lines is allowed by LSND, while
that to the right of the vertical solid line is consistent with the CDHSW
constraint; the zones above the small-dashed, medium-dashed and
big-dashed lines are allowed from Bugey, E531 and E776 respectively.\\
(b) The permiisible area from LSND is between the curved solid lines;
the areas above the small-dashed, medium-dashed and
big-dashed lines are allowed from Bugey, E531 and E776 respectively;
the region to the left of the vertical solid line is admitted from
CDHSW. \\
In each of these figures the area marked as `allowed ' is
consistent with all the constraints.\\
Figure 6: The allowed region of fig. 5c that is consistent with the
atmospheric neutrino constraint is shown shaded.\\

\newpage
\setlength{\unitlength}{1.3cm}
\begin{picture}(42.5,20.5)(-0.6,12.0)
\put(-1.0,28.0){\line(1,0){3.0}}
\put(3.0,28.0){\makebox(0,0){m$_{4}^2$}}
\put(0.5,27.5){\makebox(0,0){i~(a)}}
\put(0.5,28.4){\vector(0,1){0.2 }}
\put(0.5,28.3){\makebox(0,0){\small{$\Delta_{SOLAR}$}}}
\put(0.5,28.2){\vector(0,-1){0.2 }}
\put(-1.0,28.6){\line(1,0){3.0 }}
\put(3.0,28.6){\makebox(0,0){m$_{1}^2$}}
\put(-1.0,30.0){\line(1,0){3.0 }}
\put(3.0,30.0){\makebox(0,0){m$_{2}^2$}}
\put(0.5,30.6){\vector(0,1){0.4}}
\put(0.5,30.5){\makebox(0,0){\small{$\Delta_{ATM}$}}}
\put(0.5,30.4){\vector(0,-1){0.4}}
\put(-1.0,31.0){\line(1,0){3.0 }}
\put(3.0,31.0){\makebox(0,0){m$_{3}^2$}}
\put(-2.0,29.5){\vector(0,1){1.1 }}
\put(-2.0,29.4){\makebox(0,0){ $\Delta_{LSND}$}}
\put(-2.0,29.3){\vector(0,-1){1.1 }}

\put(8.0,28.0){\line(1,0){3.0 }}
\put(12.0,28.0){\makebox(0,0){m$_{2}^2$}}
\put(9.5,28.6){\vector(0,1){0.4}}
\put(9.5,28.5){\makebox(0,0){\small{$\Delta_{ATM}$}}}
\put(9.5,28.4){\vector(0,-1){0.4}}
\put(8.0,29.0){\line(1,0){3.0 }}
\put(12.0,29.0){\makebox(0,0){m$_{3}^2$}}
\put(8.0,30.5){\line(1,0){3.0 }}
\put(12.0,30.5){\makebox(0,0){m$_{1}^2$}}
\put(9.5,30.85){\vector(0,1){0.2}}
\put(9.5,30.75){\makebox(0,0){\small{$\Delta_{SOLAR}$}}}
\put(9.5,30.65){\vector(0,-1){0.2}}
\put(8.0,31.0){\line(1,0){3.0 }}
\put(12.0,31.0){\makebox(0,0){m$_{4}^2$}}
\put(7.0,29.725){\vector(0,1){1.125}}
\put(7.0,29.625){\makebox(0,0){ $\Delta_{LSND}$}}
\put(7.0,29,525){\vector(0,-1){1.125}}
\put(9.5,27.5){\makebox(0,0){i (b)}}

\put(-1.0,20.0){\line(1,0){3.0}}
\put(0.5,20.4){\vector(0,1){0.2}}
\put(0.5,20.3){\makebox(0,0){\small{$\Delta_{SOLAR}$}}}
\put(0.5,20.2){\vector(0,-1){0.2}}
\put(-1.0,20.6){\line(1,0){3.0}}
\put(-1.0,21.3){\line(1,0){3.0}}
\put(4.0,20.9){\vector(0,1){0.5}}
\put(4.0,20.8){\makebox(0,0){\small{$\Delta_{ATM}$}}}
\put(4.0,20.7){\vector(0,-1){0.5}}
\put(-1.0,23.1){\line(1,0){3.0}}
\put(-2.0,21.775){\vector(0,-1){1.125}}
\put(-2.0,21.875){\makebox(0,0){ $\Delta_{LSND}$}}
\put(-2.0,21.975){\vector(0,1){1.125}}
\put(3.0,20.0){\makebox(0,0){m$_{4}^2$}}
\put(3.0,20.6){\makebox(0,0){m$_{1}^2$}}
\put(3.0,21.3){\makebox(0,0){m$_{2}^2$}}
\put(3.0,23.1){\makebox(0,0){m$_{3}^2$}}
\put(0.5,19.5){\makebox(0,0){ii~(a)}}

\put(8.0,20.0){\line(1,0){3.0}}
\put(8.0,21.8){\line(1,0){3.0}}
\put(8.0,22.4){\line(1,0){3.0}}
\put(8.0,23.1){\line(1,0){3.0}}
\put(12.0,20.0){\makebox(0,0){m$_{3}^2$}}
\put(12.0,21.8){\makebox(0,0){m$_{4}^2$}}
\put(12.0,22.4){\makebox(0,0){m$_{1}^2$}}
\put(12.0,23.1){\makebox(0,0){m$_{2}^2$}}
\put(9.5,22.0){\vector(0,-1){0.2}}
\put(9.5,22.1){\makebox(0,0){\small{$\Delta_{SOLAR}$}}}
\put(9.5,22.2){\vector(0,1){0.2}}
\put(13.0,22.5){\vector(0,-1){0.5}}
\put(13.0,22.6){\makebox(0,0){\small{$\Delta_{ATM}$}}}
\put(13.0,22.7){\vector(0,1){0.5}}
\put(7.0,21.325){\vector(0,1){1.125}}
\put(7.0,21.225){\makebox(0,0){ $\Delta_{LSND}$}}
\put(7.0,21.125){\vector(0,-1){1.125}}
\put(9.5,19.5){\makebox(0,0){ii~(b)}}

\put(5.0, 17.5){\makebox(0,0){\Large Fig. 1}}

\end{picture}

\begin{thebibliography}{99}
\bibitem{fogli2} The usage of the `ratio of ratios' R as a valid
indicator of the neutrino anomaly has been recently critically
examined. see G.L. Fogli and E. Lisi, Institute for Advanced Study
report IASSNS-AST 95/21.
\bibitem{ssm2}The theoretical rates quoted are from the last one in
ref. \cite{ssm}, which is the latest in the series.
\end{thebibliography}
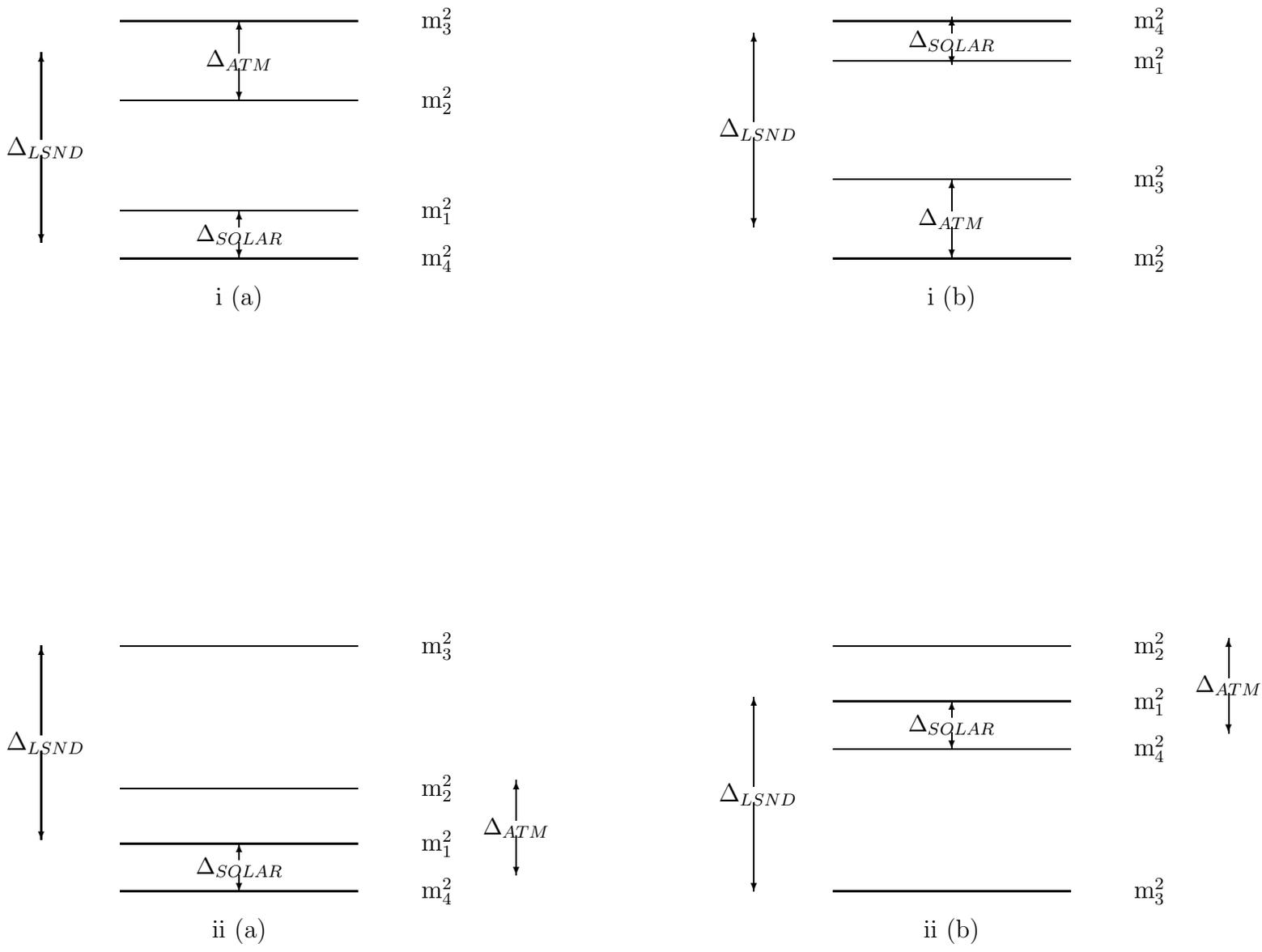
\end{document}